\documentclass[twocolumn,prd,showpacs,preprintnumbers,amsmath,amssymb]{revtex4}

\usepackage[dvips]{graphicx}
\usepackage{dcolumn}
\usepackage{amsmath}
\usepackage{amsthm} 
\usepackage{amsfonts}                        
\usepackage{amssymb}
\usepackage{bm}
\usepackage{mathrsfs}
\usepackage{bbm}

\begin{document}

  
  \providecommand{\planckl}{\ensuremath{{\lambda_{\mathbf{P}}}}}
  \providecommand{\del}{\ensuremath{\partial}}  

  \providecommand{\qm}{\ensuremath{{q^\mu}}}
  \providecommand{\qn}{\ensuremath{{q^\nu}}}
  \providecommand{\qb}{\ensuremath{{\mathbf{q}}}}
  \providecommand{\qbs}{\ensuremath{{\boldsymbol{q}}}}
  \providecommand{\kappab}{\ensuremath{{\boldsymbol{\kappa}}}}
  \providecommand{\betab}{\ensuremath{{\boldsymbol{\beta}}}}
  \providecommand{\Ub}{\ensuremath{{\mathbf{U}}}}
  \providecommand{\Ib}{\ensuremath{{\mathbf{I}}}}
  \providecommand{\Pb}{\ensuremath{{\mathbf{P}}}}
  \providecommand{\Vb}{\ensuremath{{\mathbf{V}}}}
  \providecommand{\Rb}{\ensuremath{{\mathbf{R}}}}
  \providecommand{\xb}{\ensuremath{{\mathbf{x}}}}
  \providecommand{\yb}{\ensuremath{{\mathbf{y}}}}
  
  \providecommand{\ab}{\ensuremath{{\mathbf{a}}}}
  \providecommand{\ann}{\ensuremath{\underline{a}^0}}
  \providecommand{\an}{\ensuremath{\underline{a}^\mu}}
  \providecommand{\abn}{\ensuremath{\underline{\mathbf{a}}}}
  \providecommand{\avn}{\ensuremath{\underline{a}}}
  
  \providecommand{\bb}{\ensuremath{{\mathbf{b}}}}
  \providecommand{\bnn}{\ensuremath{\underline{b}^0}}
  \providecommand{\bn}{\ensuremath{\underline{b}^\mu}}
  \providecommand{\bbn}{\ensuremath{\underline{\mathbf{b}}}}
  \providecommand{\bvn}{\ensuremath{\underline{b}}}
 
  \providecommand{\xb}{\ensuremath{{\mathbf{x}}}}
  \providecommand{\xnn}{\ensuremath{\underline{x}^0}}
  \providecommand{\xn}{\ensuremath{\underline{x}^\mu}}
  \providecommand{\xbn}{\ensuremath{\underline{\mathbf{x}}}}
  \providecommand{\xvn}{\ensuremath{\underline{x}}}  
 
  \providecommand{\kb}{\ensuremath{{\mathbf{k}}}}
  \providecommand{\knn}{\ensuremath{\underline{k}^0}}
  \providecommand{\kn}{\ensuremath{\underline{k}^\mu}}
  \providecommand{\kbn}{\ensuremath{\underline{\mathbf{k}}}}
  \providecommand{\kvn}{\ensuremath{\underline{k}}}
 
  \providecommand{\lb}{\ensuremath{{\mathbf{l}}}}
  \providecommand{\lbn}{\ensuremath{\underline{\mathbf{l}}}}
  \providecommand{\lvn}{\ensuremath{\underline{l}}}
 
  \providecommand{\pb}{\ensuremath{{\mathbf{p}}}}
  \providecommand{\pvn}{\ensuremath{\underline{p}}}
  \providecommand{\pbn}{\ensuremath{\underline{\mathbf{p}}}}

  \providecommand{\qb}{\ensuremath{{\mathbf{q}}}}
  \providecommand{\qvn}{\ensuremath{\underline{q}}}
  \providecommand{\qbn}{\ensuremath{\underline{\mathbf{q}}}}

  \providecommand{\xin}{\ensuremath{\underline{{\xi}}}} 
  \providecommand{\xibn}{\ensuremath{\boldsymbol{\underline{\xi}}}} 
 
  \providecommand{\en}{\ensuremath{\underline{{e}}_n}} 
  \providecommand{\ebn}{\ensuremath{{\underline{\mathbf{e}}_n}}}
  \providecommand{\evn}{\ensuremath{\underline{{e}}}} 
  \providecommand{\ebvn}{\ensuremath{{\underline{\mathbf{e}}}}} 

 
 \newcommand{\1}{\ensuremath{\rm \mathbbm{1}}}
  \newcommand{\wickprod}[1]{\ensuremath{\text{\bf :} #1 \text{\bf :}}}
  \newcommand{\wickdotl}{\protect{{\ :}}}
  \newcommand{\wickdotr}{\protect{{ {:}\ }}}

 
  \providecommand{\bra}[1]{\ensuremath{\left\langle {#1} \right|}}
  \providecommand{\ket}[1]{\ensuremath{\left|{#1} \right\rangle}}


\newtheorem{prop}{Proposition}
\newtheorem{defi}[prop]{Definition}
\newtheorem{thm}[prop]{Theorem}
\newtheorem{rem}[prop]{Remark}
\newtheorem{lem}[prop]{Lemma}


\title{UV Finite Field Theories on Noncommutative Spacetimes:
\\ the Quantum Wick Product and Time Independent Perturbation Theory}

\author{Marcel Kossow}

 \email{mkossow@physnet.uni-hamburg.de}
 \affiliation{ I. Institut f\"ur Theoretische Physik, Universit\"at Hamburg, %
              Jungiusstra\ss e 9,\\ D - 20355 Hamburg, Germany}



\date{\today}


\begin{abstract}
In this article an energy correction is calculated in the time independent 
perturbation setup
using a  regularised ultraviolet finite Hamiltonian on the  noncommutative 
Minkowski space.
The correction to the energy is invariant under rotation and translation  
but is not Lorentz covariant and
this leads to a distortion of the dispersion relation.  
In the limit where the  noncommutativity  vanishes the common 
quantum field theory on the commutative Minkowski space is reobtained.
\end{abstract}



\maketitle

\section{\label{sec:introduction}Introduction}

Spacetime uncertainties were considered earlier as a possible way to regularise
 the divergencies of a 
point interaction \cite{Snyder:1946qz}. Taking the  gedankenexperiment into 
account that the principles 
of classical general 
relativity and of quantum mechanics  lead to spacetime uncertainties  
a regularising effect can be argued to appear at the Planck length 
$\planckl=\sqrt{G\hbar/c^3}\approx 1.6\cdot 10^{-35}\rm m$ 
 \cite{Doplicher:2001qt}. 
In \cite{Doplicher:1995tu} the spacetime uncertainties 
\begin{eqnarray*}
 \Delta x_0(\Delta x_1\Delta x_2+\Delta x_3)&\gtrsim&\planckl^2\ \text{and}\\
 \Delta x_1\Delta x_2+\Delta x_2 \Delta x_3 + \Delta x_3 \Delta x_1&\gtrsim&
\planckl^2\ ,
\end{eqnarray*}
were found, and can be implemented in a Poincar\'e invariant manner by 
appropriate commutation 
relations of the spacetime coordinates considered as  noncommuting unbounded, 
self-adjoint operators $q^\mu$:
\begin{eqnarray}\label{quantumconditions}
 \   [q^\mu,q^\nu]&=&i\planckl^2 Q^{\mu\nu}\\
 \  [q^\mu,Q^{\nu\rho}]&=&0\label{qmuqmunu=0}\\
 \  Q_{\mu\nu} Q^{\mu\nu}&=&0\label{qmunu=0}\\
 \ (\epsilon^{\mu\nu\lambda\rho}q_\mu q_\nu q_\lambda q_\rho)^2&=&\ \planckl^8
\  \1\label{lambda^8}
\end{eqnarray}
The commutators have to be central \eqref{qmuqmunu=0} and 
condition \eqref{qmunu=0} and \eqref{lambda^8} fix a 
topological manifold $\Sigma\sim TS^2\times\{1,-1\}$, the joint spectrum of 
the $Q^{\mu\nu}$. 
\ 

In the following we set $\planckl=c=\hbar=1$. 
Inspired by the algebraic approach to quantum mechanics, 
the noncommutative Minkowski space $\mathcal{E}$ is constructed  as a
 noncommutative 
$C^*$-algebra generated by the Weyl symbols of the spacetime coordinates
 \cite{Doplicher:1995tu}. 
$\mathcal{E}$ is shown to be isomorphic to 
$\mathcal{C}_0(\Sigma,\mathcal{K})$; 
the continuous functions vanishing at infinity and taking values in the compact
 operators 
$\mathcal{K}$  on a separable, infinite dimensional Hilbert space.  
This is in  analogy with the commutative framework, where one has a commutative
 $C^*$-algebra  
of continuous functions vanishing at infinity and taking 
values in $\mathcal{K}$ on the commutative Minkowski space.
\

In \cite{Doplicher:1995tu} a 
quantum field theory  which is fully 
Lorentz covariant is formulated on  the noncommutative Minkowski space.
 In order to apply a unitary perturbation theory \cite{Bahns:2002vm} 
there are two inequivalent approaches
which can be taken \cite{Bahns:2002fc}. The first uses the perturbation setup
 according to Dyson.
An effective Hamiltonian with a nonlocal kernel is defined, averaging the 
noncommutativity
at each vertex. A modification to this Hamilton approach  replaces the limit of
 coinciding points
by a suitable generalisation and this  yields an $S$-Matrix, which has the 
property of being 
ultraviolet finite, term by term. 
The second approach, which is not 
considered in this article, uses the Yang-Feldman equation, where the quantum
 fields are
treated as $q$-distributions and products of fields called quasi planar Wick
 products 
are defined by considering only $q$-local counter-terms \cite{Bahns:2004fc}.
\  

In this article we only consider the ultraviolet finite Hamilton 
approach using the regularised 
Wick monomials on the noncommutative Minkowski space  also 
called the quantum Wick product \cite{Bahns:2003vb}. The UV-finiteness is a 
comfortable feature of this theory, since in other approaches to 
noncommutative field theory
there appear serious UV and UV/IR mixed divergencies \cite{Minwalla:1999px}, 
which 
need strong efforts and new concepts to handle with, for example in the 
Yang-Feldman approach
\cite{Bahns:2004fc} but also in NC-QFT on the 2$n$ quantum plane  
\cite{Rivasseau:2005bh} and
\cite{Grosse:2006tc}. The existence of the adiabatic limit in the theory 
of the quantum  Wick product remains 
an open question, but we show in this article that it exist at least 
in a $\phi^3$ theory up to second order time independent perturbation theory.
\ 

The starting point is a suitable  definition of the Wick ordered product of 
 fields on the
commutative Minkowski space, which also works in curved spacetime i.e. $n=2$:
\begin{eqnarray*}
 \wickprod{\phi^2(x)}\ &=& \lim_{y\to x} (\phi(x)\phi(y)-\bra{\Omega}\phi(x)
\phi(y)\ket{\Omega} )\ .
\end{eqnarray*}
In the limit of coinciding events on a commutative spacetime, 
the right hand side  yields a well defined distribution and is equivalent to  
putting all creation
operators on the left. However on the noncommutative Minkowski space it is not 
possible
to perform this limit. This situation is comparable to the one in quantum 
mechanics.
In the  quantum mechanical phase space it also makes no sense
taking the limit of coinciding points. Instead, one  would evaluate the 
differences
in coherent states to minimise the distance. 
From this point of view one can introduce mean and relative coordinates
on the noncommutative Minkowski space and replace the limit of coinciding 
points  by
a suitable  generalisation, the quantum diagonal map $E^{(n)}$  
\cite{Bahns:2003vb}. 
This map  evaluates   
the relative coordinates in pure states and
restricts the mean on the sub-manifold $\Sigma_1\subset\Sigma$, which can be  
shown to be 
homeomorphic to $S^2\times \{1,-1\} $. The result is a regularised nonlocal 
 version of
the Wick monomial $\wickprod{\phi_R^{(n)}(\tilde{q})}$. 
While depending solely on the mean coordinate $\tilde{q}$, the regularised Wick
 monomial is
a constant function in $\Sigma_1$ and transforms covariantly under rotation and
 translation but
not under Lorentz boosts. The quantum diagonal map behaves as follows 
\begin{eqnarray*}
 \wickprod{\phi_R^{(n)}(\tilde{q})} \ &=& E^{(n)}(\wickprod{{\phi(q_1)\dots
\phi(q_n)}})\nonumber\\
    &=& \int_{\mathbb{R}^{4n}} dk_1\dots dk_n\  r_n(k_1\dots k_n)\\\nonumber
   &&\quad \times\  \wickprod{\check{\phi}(k_1)\dots \check{\phi}(k_n)}\ 
 e^{i(k_1+\dots+k_n)\tilde{q}}\ .
\end{eqnarray*}
The kernel $r_n(k_1,\dots,k_n)$ is then the Fourier transform of the kernel 
$\tilde r_n$ given below. 
\

Since $\int_{q^0=t}d\qb$ is a positive centre valued functional on 
(the multiplier algebra of) $\mathcal{E}$ \cite{Doplicher:1995tu},
 one can take the symbol $\wickprod{\phi_R^{(n)}(x)}$ of  
$\wickprod{\phi_R^{(n)}(\tilde{q})}$, which is given by:
\begin{eqnarray*}
  \wickprod{\phi_R^{(n)}(x)}&=&\int_{\mathbb{R}^{4n}}dk_1\dots dk_n\ r_n(k_1,
\dots,k_n)\\ 
  &&\quad\quad\times\ \wickprod{\check{\phi}(k_1)\dots \check{\phi}(k_n)}\ 
             e^{i(k_1+ \dots + k_n)x}\nonumber\\
 &=& c_n \int_{\mathbb{R}^{4n}} da_1\dots da_n\  \tilde{r}_n(x-a_1,
\dots,x-a_n)\\\nonumber 
 && \quad\quad   \times\   \wickprod{\phi(a_1)\dots\phi(a_n)}\ .
\end{eqnarray*}
$c_n=\tfrac{n^2}{(2\pi)^{2(n-2)}}$ is a constant, depending only on the power
 of the Wick monomial. 
The nonlocal kernel $\tilde{r}_n$ is calculated to: 
\begin{eqnarray*}
  \tilde{r}_n(a_1,\dots,a_n)= e^{-\tfrac{1}{2}|a_1|^2-\dots-\tfrac{1}{2}
       |a_n|^2 } \  \delta^{(4)}(a_1+\dots+a_n)\ .    
\end{eqnarray*} 
The Gaussian functions, decreasing with  the 
noncommutativity parameter (or Planck length),
are due to the evaluation of the relative coordinates in best localised states 
and the Dirac delta function respects the fact that we do not evaluate the 
mean in pure states.
The interaction Hamiltonian can  be  defined by: 
\begin{eqnarray*}
  H_I (t):= {\lambda} \int_{\tilde{q}^0=t} d^3\tilde{q}\  \mathcal{L}_{\rm eff}
(\tilde{q})
\end{eqnarray*}
where the effective Lagrangian of the mean is given by 
\begin{eqnarray*}
  \mathcal{L}_{\rm eff}(\tilde{q}) &=&\tfrac{1}{n!}\ 
  \wickprod{\phi^{(n)}_R(\tilde{q})}\\\nonumber
 &=&\int dk_1\dots\ dk_n\  \check{\mathcal{L}}_{\rm eff}(k_1,\dots k_n)\  
    e^{i(k_1+\dots+k_n)\tilde{q}}
\end{eqnarray*}
and the effective Lagrangian in momentum space is
\begin{eqnarray*}
   \check{\mathcal{L}}_{\rm eff}(k_1,\dots,k_n)=\tfrac{1}{n!}\
    \wickprod{\check{\phi}(k_1)\dots \check{\phi}(k_n)}\ .
\end{eqnarray*}
Again  the symbol $\mathcal{L}_{\rm eff}(x)$ of 
$\mathcal{L}_{\rm eff}(\tilde{q})$ is taken
and the effective interaction Hamiltonian is defined by  
$$
H_I^\lambda := \int dx\ \delta(t-x^0)\ \lambda(x)\ 
\mathcal{L}_{\rm eff}(x)\ \ . 
$$
The coupling constant $\lambda$ is turned  into 
a Schwartz function $\lambda(\cdot)\in\mathcal{S}(\mathbb{R}^4)$ and acts as 
an adiabatic switch 
to regularise the infrared regime.
It has been  shown that this regularised interaction  yields a formal Dyson 
series, which is ultraviolet finite term by term \cite{Bahns:2003vb} and for 
a brief discussion on this see \cite{Bahns:2004mm}.
\

It should be  mentioned that this approach differs sensitively from the 
approach of 
smeared field operators  \cite{Denk:2004pk} in the 
context of general nonlocal kernels.
This is due to the fact that the 
quantum diagonal map smears out only the relative coordinates, which are at 
least $n-1$ coordinates.
Thus one coordinate, the mean, is left un-smeared and therefore 
this theory is local in the (symbol of the) mean. 
In fact, the theory would be nothing else but  a theory with Wick ordered 
products of fields, 
where first the fields are smeared with a Gaussian function (decreasing with 
the Planck length),  iff
we would evaluate the mean coordinate in pure states, too. The discussion on  
ultraviolet and infrared  divergences  as well as  renormalisation then 
reduces to  one with smeared field operators.
However, since  $\int_{q^0=t}d\qb$ is a positive 
trace, there is at first glance no need to evaluate the mean in pure states 
and thus we preserve locality  in the mean. 
Unfortunately this results in having  to deal with serious divergencies in 
the adiabatic limit \cite{Bahns:2004mm} in
the framework of time dependent perturbation theory according to Dyson. 
Furthermore  it is not clear whether a (mass-)renormalisation can be 
performed in analogy to the standard renormalisation
procedure in the commutative quantum field theory due to the acausality of 
the theory -- explicitly the acausality of the generalised propagator 
\cite{Piacitelli:2004rm}. 
The way the generalised propagator depends on the time variable -- in the 
case of the regularised field monomials -- causes the most serious 
difficulties. Therefore  our motivation is to apply the time independent 
perturbation setup in the ultraviolet finite Hamilton approach.

\section{\label{spectrum} Time Independent Perturbation Setup}
\begin{figure}[t]
  \begin{center}
   \includegraphics[width=0.48\textwidth]{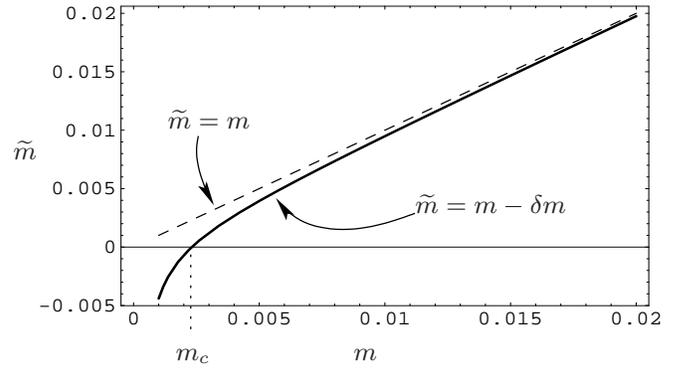}
   \caption{The physical mass vs. bare mass in 
            units of the noncommutative scale 
         $\lambda_{\rm nc}$ (or Planck length \planckl). 
         The graph is plotted with {\tt Mathematica 5.0} for
         $m\in [0.001,0.02]\cdot\lambda_{\rm nc}^{-1}$. The  straight line 
         corresponds to $\widetilde m=m-\delta m$ and the dashed line to 
         $\widetilde m=m$.}
    \label{fig:mtilde}
  \end{center}
\end{figure}

We use the time independent perturbation theory  in the formulation of 
Rayleigh-Schr\"odinger \cite{Reedsimon}
and calculate the energy correction in the vacuum and the improper 
one-particle state  to second order.  
The infrared divergent part of the  expectation value in the one-particle 
state then precisely cancels  with the divergent expectation value in the 
vacuum state. We define the formal Rayleigh-Schr\"odinger series by
$$
 E(\lambda):=E_0+\sum_{n=1}^\infty \alpha_n^{\lambda}\ ,
$$
where we switched the coupling constant $\lambda$ into a Schwartz function 
$\lambda(\cdot )\in\mathcal{S}(\mathbb{R}^4)$ (adiabatic switch) 
to regularise  the infrared regime. At later stage  we  perform the 
 adiabatic limit. From now on $q$ denotes the four-momentum. The first order 
correction to the energy is zero due to normal ordering 
($\alpha^\lambda_1[\cdot]=0$).  The dot denotes the evaluation 
either in the vacuum or in the one-particle state. The formal coefficients to 
second order $\alpha_2^\lambda[\cdot] $ are given by:
\begin{eqnarray*}
 \alpha_2^\lambda[\Omega] &=& -\bra{\Omega}H_I^\lambda\ 
H_0^{-1}H_I^\lambda\ket{\Omega}\ \text{and}\\
\bra{q'}q\rangle \ \alpha^\lambda_2[q] &=& -\bra{q'}
H^\lambda_I\ (H_0-\omega_\qb)^{-1}H^\lambda_I\ket{q}\ .
\end{eqnarray*}
Therein  we  normalised the coefficient with the delta function, since we 
deal with improper momentum states $\ket{q}$. 
$\ket{\Omega}$ denotes the Fock vacuum.
The free Hamiltonian  $H_0$ acts on the improper one-particle state by
$$
 H_0 \ket{p}={\omega_\pb}\ket{p}\quad\text{and }\quad \omega_\pb=
\sqrt{\pb^2+m^2}
$$
and the interaction Hamiltonian at $t=0$ is given by:
\begin{eqnarray}\label{Hphi0}
 H^\lambda_I =\tfrac{1}{n!}\int d\xb\ \lambda(\xb)
\wickprod{\phi_R^{(n)}(0,\xb)}\ .
\end{eqnarray}
It should be  mentioned that taking the regularised Wick monomials at $t=0$ 
differs from taking the time-zero-fields. In fact the time-evolution 
of $H_I^\lambda$ is given by the expression \cite{Bahns:2004mm}: 
\begin{eqnarray*}
 H_I^\lambda (t) &=& e^{iH_0t}\  \int \big[\delta(x^0-\tau) \lambda(x)\ 
                      \mathcal{L}_{\rm eff}(x)\big]_{\tau=0} dx\  e^{-iH_0t}\\
                  &=&\tfrac{c_n}{n!}\int d\xb \int da_1\dots da_n\ 
\lambda(0,\xb)\ \\
                  &&\quad\quad\quad\times\ \tilde{r}_n((0,\xb)-a_1,\dots,
(0,\xb)-a_n)\\
                 &&\quad\quad\quad\times\   e^{iH_0t}\wickprod{\phi(a_1)\dots
\phi(a_n)}e^{-iH_0t}\ .
\end{eqnarray*}
Before we start our computations the following convention is adopted: 
the free spin-zero fields are defined by
$$
\phi(x) =\frac{1}{(2\pi)^{3/2}}\int \frac{d\kb}{\sqrt{2\omega_\kb}} 
\Big[e^{ikx}a_\kb^++e^{-ikx}a_\kb \Big]
$$ 
and the Fock states by
$$
\ket{k_1\dots k_n}=\frac{1}{\sqrt{n!}}a_{\kb_1}^+
\dots  a_{\kb_n}^+\ket{\Omega}\ ,\ \  a_{\kb_i}\ket{\Omega}=0\ ,
$$
$$
\quad[a_{\kb'},a_{\kb}^+]=\delta^{(3)}(\kb'-\kb)
$$
such that the  $n$-particle state is normalised to
$$
 \bra{ k_1\dots k_n}{ p_1\dots p_n}\rangle = \frac{1}{n!} \sum_\pi\ 
\prod_{i=1}^n \delta^{(3)}(\kb_i-\pb_{\pi(i)})\ .
$$
The energy correction in the vacuum state is then:
\begin{eqnarray}\label{eq:Vakuumkorrektur1}
  \alpha^\lambda_2 [\Omega]&=&\frac{1}{n!^2}\ \int 
  \frac{d\pb_1\dots d\pb_n}{\omega_{\pb_1}+\dots+\omega_{\pb_n}}
\Big[\nonumber \\
&& \int d\xb d\yb\ \lambda(\xb)\  \lambda(\yb)\\\nonumber 
         &&\ \times\  \bra{\Omega}\wickprod{\phi_R^{(n)}(0,\xb)}\ket{p_1\dots
 p_n}\\\nonumber
       &&\ \times\ \bra{p_1\dots p_n}\wickprod{\phi_R^{(n)}(0,\yb)} 
\ket{\Omega}\Big]\ .
\end{eqnarray}
The energy correction in the improper one-particle state consists of two terms:
\begin{eqnarray}\label{eq:alphaq}
 \alpha^\lambda_2[q]=  B^\lambda_2[q]+ C^\lambda_2[q]\ .
\end{eqnarray}
The first term is
\begin{eqnarray*}
\bra{q'}q\rangle\  B^\lambda_2[q]&=&\frac{1}{n!^2}\int\frac{d\pb_1\dots 
d\pb_{n-1}}{\omega_{\pb_1}+\dots+
       \omega_{\pb_{n-1}}-\omega_\qb}\ \Big[ \nonumber \\
&& \int d\xb d\yb\  \lambda(\xb)\  \lambda(\yb)\ \\\nonumber
&&\times\  \bra{q'}\wickprod{\phi_R^{(n)}(0,\xb)}\ket{p_1\dots 
p_{n-1}}\\\nonumber
&&\times\ \bra{p_1\dots p_{n-1}}\wickprod{\phi_R^{(n)}(0,\yb)} \ket{q}\Big]
\end{eqnarray*}
and the second is
\begin{eqnarray*}
\bra{q'}q\rangle\  C^\lambda_2[q]&=&\frac{1}{n!^2}\int \frac{d\pb_1\dots  
d\pb_{n+1}}{\omega_{\pb_1}+
       \dots +\omega_{\pb_{n+1}}-\omega_\qb}\Big[\nonumber \\
&& \int d\xb d\yb\  \lambda(\xb)\  \lambda(\yb)\  \\\nonumber
&&\  \times\ \bra{q'}\wickprod{\phi_R^{(n)}(0,\xb)}
\ket{p_1\dots p_{n+1}} \\\nonumber
&&\ \times\     \bra{p_1\dots p_{n+1}}\wickprod{\phi_R^{(n)}(0,\yb)} 
\ket{q}\Big]\ .
\end{eqnarray*}
Now we are ready to define the energy correction.
\begin{defi} At second order time-independent perturbation theory 
the renormalised energy correction is defined in the adiabatic limit
$$ 
\delta E(\qb):=\lim_{\lambda(\cdot)\to 1} \alpha^\lambda_2[q]-
\alpha^\lambda_2[\Omega] \ ,
$$
such that the effective particle energy up to second order 
is given by $E(\qb)=\omega(\qb)-\delta E(\qb)$.
\end{defi}
In the following we  study this correction in more detail.
The evaluation of the coefficients $\alpha^\lambda_2$ uses techniques similar 
to \cite[appendix]{Bahns:2003vb} and in order 
to keep the following formulae simple,  we restrict our discussion  to a 
$\wickprod{\phi_R^{(3)}}$-theory; 
the general case $\wickprod{\phi_R^{(n)}}$ can be performed analogously. 
As an example  the correction in the vacuum state is calculated in the 
appendix.
The correction to the improper one-particle state can be calculated  
analogously.
We obtain an expression for $\alpha^\lambda_2[\Omega]$, 
which diverges in the adiabatic limit $\lambda(\xb)\to 1$. Take for example 
the cut-off function $\lambda(\xb)$ as a Gaussian function 
$\lambda(\xb)=\exp\{-|\xb|^2/\alpha\}$
with the dumping parameter $\alpha$ and perform the limit $\alpha\to\infty$. 
 In Fourier space this would turn $\check\lambda(\pb)$ into a delta-function.
Thus in the adiabatic limit this expression is not well-defined due to the 
appearance of the square of a delta-function 
$|\check\lambda(\pb)|^2\to |\delta(\pb)|^2$ \eqref{eq:VakuumskorrekturA}:
\begin{eqnarray}\label{eq:Vakuumskorrektur2}
  \alpha^\lambda_2[\Omega]
&=&\frac{1}{(2\pi)^9 3^4 16}\int \frac{d\pb_1 d\pb_2 d\pb_3}{\omega_{\pb_1}+
  \omega_{\pb_2}+\omega_{\pb_3}}\nonumber\Big[\\\nonumber
&&\times\ |\check{\lambda}(\tfrac{1}{3}( \pb_1+\pb_2+\pb_3))|^2 
   \ \frac{1}{\omega_{\pb_1} \omega_{\pb_2}\omega_{\pb_3}} \\
&&\times\   \exp\Big\{\sum_{i=1}^3 |p_i-\tfrac{1}{3}\sum_{j=1}^3 p_j|^2\Big\}
\Big]\ .\nonumber
\end{eqnarray}
In the first term of the energy correction \eqref{eq:alphaq} in the improper 
one-particle 
state, $B^\lambda_2[q]$, we separate the one-particle state, 
take  the adiabatic limit $\lambda(\xb)\to 1$ and then switch to 
momentum space: 
\begin{eqnarray}\label{eq:einteilchenkorrektur2}
\langle q'|q \rangle\  B^\lambda_2[q]&=&\langle q'|q \rangle 
\frac{1}{(2\pi)^6 16}  
    \int d\pb_1 \Big[\nonumber \\\nonumber
&& \frac{ e^{-|\pb_{1}|^2 } 
             e^{-|\pb_{1}-\qb|^2 } 
             e^{-|\qb|^2 } }{(\omega_{\pb_1}+\omega_{\pb_1-\qb}-\omega_{\qb})
              \omega_{\pb_1}\omega_{\pb_1-\qb}\omega_{\qb}} \\\nonumber
&&\times\ e^{-|\omega_{\pb_{1}}-\frac{1}{3}(\omega_{\pb_{1}}+
\omega_{\pb_{1}-\qb}-\omega_{\qb})|^2}
\\\nonumber
 &&\times\  e^{-|\omega_{\pb_{1}-\qb}-\frac{1}{3}(\omega_{\pb_{1}}+
\omega_{\pb_{1}-\qb}-\omega_{\qb})|^2}
\\\nonumber
 &&\times\ e^{-|-\omega_{\qb}-\frac{1}{3}(\omega_{\pb_{1}}+
\omega_{\pb_{1}-\qb}-\omega_{\qb})|^2}\Big]\ .
\\\nonumber
\end{eqnarray}
The second term $C^\lambda_2[q]$ in \eqref{eq:alphaq} is given by the sum:
\begin{eqnarray}\label{eq:clambda}
C^\lambda_2[q]= D^\lambda_2[q]+ E^\lambda_2[q]\ .
\end{eqnarray}
In the first term  $D^\lambda_2[q]$ of \eqref{eq:clambda} 
we also separate the one-particle state and take the adiabatic limit:
\begin{eqnarray*}
\langle q'|q\rangle\ D^\lambda_2[q]&=&\langle q'|q\rangle  \frac{1}{(2\pi)^616}
\int d\pb_1\ \Big[ \\\nonumber 
&& \frac{e^{-|\pb_{1}|^2 } e^{-|\pb_{1}+\qb|^2 } e^{-|\qb|^2 }}{(
                       \omega_{\pb_1}+\omega_{\pb_1+\qb}+\omega_{\qb})
                       \omega_{\pb_1}\omega_{\pb_1+\qb}\omega_{\qb}}\\\nonumber
&&\times\ e^{-|\omega_{\pb_1}-\frac{1}{3}(\omega_{\pb_{1}}+
\omega_{\pb_{1}+\qb}+\omega_{\qb})|^2}
\\\nonumber
&&\times\ e^{-|\omega_{\pb_{1}+\qb}-\frac{1}{3}(\omega_{\pb_{1}}+
\omega_{\pb_{1}+\qb}+\omega_{\qb})|^2}
\\\nonumber
&&\times\ e^{-|\omega_{\qb'}-\frac{1}{3}(\omega_{\pb_1}+
\omega_{\pb_{1}+\qb}+\omega_{\qb})|^2}\Big]
\nonumber\ .
\end{eqnarray*}
In the second term  $E^\lambda_2[q]$ in \eqref{eq:clambda} we also separate 
the delta-function 
and obtain an expression, which is not well-defined  for $\lambda(\xb) \to 1$. 
Fortunately $E^\lambda_2[q]$ coincides exactly with 
the vacuum expectation value $\alpha^\lambda_2[\Omega]$, since we normalised it
with the delta function: 
\begin{eqnarray*}
\langle q'|q \rangle\ E^\lambda_2[q]&=&\langle q'|q \rangle 
\frac{1}{(2\pi)^93^4 16} 
\int \frac{d\pb_1 d\pb_2 d\pb_3}{\omega_{\pb_1}+
       \omega_{\pb_2}+\omega_{\pb_3}}\Big[ \nonumber\\
&& \frac{1}{\omega_{\pb_1}\omega_{\pb_2}\omega_{\pb_3}} 
  |\check{\lambda}(\tfrac{1}{3}(\pb_1+\pb_2+\pb_3))|^2\\  \nonumber
&&\times\   \exp\Big\{\sum_{i=1}^3 |p_i-\tfrac{1}{3}\sum_{j=1}^3 p_j|^2\Big\}
\Big]\nonumber\ .
\end{eqnarray*}
The renormalised correction to the energy is then given by:
\begin{eqnarray}\label{de=b+e}
\delta E(\qb) &=&\lim_{\lambda(x)\to 1}B^\lambda_2[q]+D^\lambda_2[q]
\end{eqnarray}
The adiabatic limit has to be understood in the sense of distributions,
 since the
integral kernels of both $B^\lambda_2[q]$ and $D^\lambda_2[q]$ remain 
Schwartz functions 
 after the adiabatic limit is carried out, which is shown in the following 
theorem.
\begin{thm}\label{prop:deltae}
\begin{itemize}
\item[$(i)$] In the ultraviolet finite Hamilton approach \cite{Bahns:2003vb}
             the energy correction $\delta E(\qb)$ up to second order time 
independent 
             perturbation theory is finite in the adiabatic limit 
             for  $\qb\in \mathbb{R}^3$ and  mass  $m>0$,
             contrary to the quantum field theory on the commutative
 Minkowski space.
\item[$(ii)$]  $\delta E(\qb)$ is invariant under rotation and translation 
but not Lorentz covariant.
\item[$(iii)$] In the limit  where the noncommutativity vanishes 
$\planckl\to 0$, we obtain the 
               massive  quantum field theory on the commutative Minkowski 
space after the
                introduction of a cut-off function. 
\end{itemize}
\end{thm}
It is convenient to introduce spherical coordinates $q:=|\qb|$, $p:=|\pb|$, 
$\theta$, $\varphi$ and 
$\omega_{p,q,\theta}=\sqrt{q^2+p^2+2qp\cos\theta+m^2}$. The energy
 correction \eqref{de=b+e}
is then given by:
\begin{eqnarray}\label{deltaeqkugel}
\delta E(q)&=&\frac{e^{-\frac{10}{3}q^2} e^{-2m}}{16(2\pi)^5{\omega_q}} 
                 \int_0^\infty dp\ \int_{-1}^1 
           d\cos\theta\  \Bigg[\nonumber\\
        &&\times\  p^2\ \frac{ e^{-\frac{10}{3}p^2 } 
e^{-\frac{8}{3}qp\cos\theta} 
             e^{-\omega_p \omega_{p,q,\theta}+\omega_p\omega_q}}
           {\omega_{p}\omega_{p,q,\theta}} \\
&&\times\  \Bigg( \frac{e^{\omega_{p,q,\theta}\omega_q}}{{\omega_{p}+
\omega_{p,q,\theta}-\omega_{q}}}\ 
      +\ \frac{e^{-\omega_{p,q,\theta}\omega_q}}{\omega_{p}+
\omega_{p,q,\theta}+\omega_{q}}\Bigg) \Bigg]
\nonumber\ .  
\end{eqnarray}
\begin{proof}
$(i)$: In our framework the energy correction of the  commutative 
$\phi^3$-theory 
$\delta \tilde{E}(\qb)$ is given by:
\begin{eqnarray}\label{kle}
\delta \tilde{E}(\qb)&=& \lim_{\Lambda\to\infty} \frac{1}{(2\pi)^6}\frac{1}{8}
\int_{-\Lambda}^\Lambda\frac{d\pb_1}{\omega_{\pb_1}\omega_{\pb_1+\qb}
\omega_{\qb}}\\\nonumber
&&\quad\quad\quad \times\ \frac{\omega_{\pb_1}+
\omega_{\pb_1+\qb}}{(\omega_{\pb_1}+\omega_{\pb_1+\qb})^2-\omega_{\qb}^2}\ .
\end{eqnarray}
This integral diverges obviously for $\Lambda\to\infty$ logarithmically. 
Contrary to this we show that  $\delta E(\qb)$ is a well-defined
continuous function in $\qb$, which is bounded on  $\mathbb{R}^3$ and 
 vanishes in the limit 
 $|\qb|\to\infty$. 
\ 
 
First we observe that the integrals in  \eqref{deltaeqkugel} are 
well-defined.
Therefore we estimate the denominator in the first term in
 \eqref{deltaeqkugel}: 
$$
\frac{1}{\omega_{p}+\omega_{p,q,\theta}-\omega_{q}}
           \geq \frac{\omega_{p}+\omega_{p,q,\theta}+\omega_{q}}{m^2}
$$
for all  $p\in\mathbb{R}$, $\theta\in [0,\pi]$, fixed $q\in\mathbb{R}$ and  
$m>0$.
The fact that $(\omega_{p}+\omega_{p,q,\theta}-\omega_{q})$ gets  arbitrarily
 close to zero for 
$p\to\infty$ is not problematic since the remaining factors form a Schwartz 
function in $p$ 
for $\theta\in [0,\pi]$, fixed $q\in\mathbb{R}$ and  $m>0$ . 
We now estimate the exponential functions of \eqref{deltaeqkugel} for 
$q,p\gg 0$:
$$
 e^{-\frac{8}{3}qp\cos\theta} 
 e^{-\omega_p \omega_{p,q,\theta}+\omega_p\omega_q} e^{\omega_{p,q,\theta}
\omega_q}
\leq e^{\frac{4}{3}(p^2+q^2)}\ .
$$
Replacing the expression on the left hand side by the right hand side  
proves that the integral over the
first term in \eqref{deltaeqkugel} is well-defined.  
It follows then that the second term is also well-defined.
Integrating over $\cos\theta$ and $p$ yields  a continuous function in $q$
 which is majorised by 
$\exp\{\tfrac{4}{3}q^2\}$.
Together  with  the pre-factor $\exp\{-\tfrac{10}{3}q^2-2m\}(q^2+m^2)^{-1/2}$
 we obtain a 
function which is bounded on $\mathbb{R}^3$ and vanishes for $q\to\infty$, 
 for any  $m>0$.\\
$(ii)$: $\delta E(\qb)$ is obviously invariant under rotation and translation,
 but  the 
        Gaussian factors fail to be Lorentz covariant. \\
$(iii)$: The noncommutative parameter $\planckl$ appears only 
         in the exponents. In the case $\planckl\to 0$ they tend to 1 and we 
obtain \eqref{kle}.
\end{proof}
It is now  convenient to introduce the shift in the mass as the value of the 
energy correction
at zero momentum.
\begin{defi} 
The renormalised  mass correction up to second order time independent 
perturbation theory $\delta m$
is defined as the limit
$$\delta m:= \lim_{q\to 0}\delta E(q)\quad,
$$
such that the effective (physical) mass is given by $\widetilde{m}=m-\delta m$.
\end{defi}
Note, in this context we do not have an infinite bare mass contrary 
to the commutative field theory. 
From  \eqref{deltaeqkugel} it
follows:
\begin{eqnarray}\label{massenkorrektur2}
\delta m &=&  \frac{1}{(2\pi)^5} \frac{1}{8} \frac{\ e^{-2m}}{m} 
         \int_0^\infty dp\  
           \frac{p^2e^{-\frac{13}{3}p^2}}{p^2+m^2}\Bigg[\\\nonumber
&&\ \frac{e^{2\sqrt{p^2+m^2} m}}{2\sqrt{p^2+m^2}-m}  +
\ \frac{1}{2\sqrt{p^2+m^2}+m}\ \Bigg] 
\end{eqnarray}
and we see  from theorem \ref{prop:deltae} that:
\begin{itemize}
\item[$(i)$] Contrary to the commutative  case the correction to the  mass 
            $\delta m$ is finite for any $m>0$.
\item[$(ii)$] In the limit where the noncommutativity vanishes 
($\planckl\to 0$) we receive 
              the correction to the mass of the commutative quantum field 
theory.
\end{itemize}
In FIG. \ref{fig:mtilde} the physical mass is plotted as a function
of the bare mass in units of the noncommutativity scale. The dashed line 
corresponds to the unrenormalised mass $\widetilde m=m$
and the straight line to the renormalised physical mass 
$\widetilde m=m-\delta m$.
What we see is that for masses at the noncommutativity scale 
($\widetilde m>0.01\cdot m_{\rm nc}$)
the physical mass is equal to the bare mass. If we assume that there exists
 no negative physical
mass then we have  a lower limit for the bare mass  
$m_{\rm c}\approx 2.317\cdot 10^{-3}\cdot m_{\rm nc}$
at the point where the physical mass is zero ($\widetilde m=0$).
\ 
\begin{figure}[t]
  \begin{center}
   \includegraphics[width=0.48\textwidth]{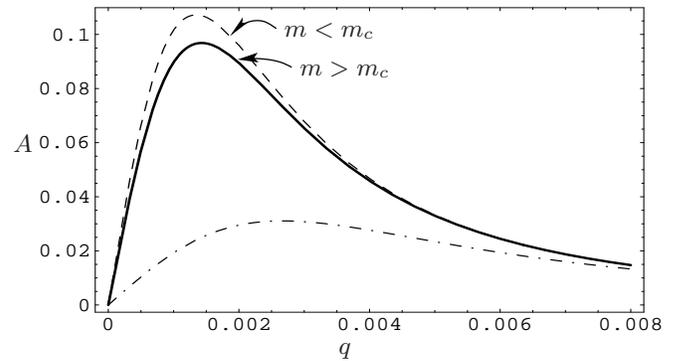}
   \caption{Radial deviation $A(q)=v(q)-\tilde{v}(q)$ of the group-velocity in 
            units of the noncommutative scale 
         $\lambda_{\rm nc}$ (or Planck length \planckl). The graph is plotted 
         with {\tt Mathematica 5.0} for
         $q\in [0,0.008]$. The  straight line corresponds to 
         $m=2.4\cdot 10^{-3}\cdot\lambda_{\rm nc}^{-1}$, 
         the  dashed line to $m=2.3\cdot 10^{-3}\cdot\lambda_{\rm nc}^{-1}$ 
         and the
       dashed-dotted line to $ m=4.0\cdot 10^{-3}\cdot\lambda_{\rm nc}^{-1} $.}
    \label{fig:verzerrung}
  \end{center}
\end{figure}
\ 
The group-velocity can also be calculated:
$$\mathbf{v} (\qb) = \nabla_\qb E(\qb)\ .
$$
The energy correction up to second order is  $E(\qb)=\omega_\qb-\delta E(\qb)$.
Obviously $\mathbf{v}(\qb)$ is again covariant under rotation and translation,
 but not covariant under  
Lorentz transformation. Therefore  we take again spherical coordinates 
$q:=|\qb|,\theta,\varphi$ and use the
fact that $|\mathbf{v}(\qb)|= v(q)$:
\begin{eqnarray}\label{vgruppe}
 v(q)= \frac{q}{\sqrt{q^2+m^2}}-\frac{\partial}{\partial q}\delta E(q)\ .
\end{eqnarray}
In doing so,  $v(q)$ can be compared with the Lorentzian invariant 
group velocity corresponding to the  renormalised mass: 
$$
\tilde{v}(q):=\frac{q}{\sqrt{q^2+(m-\delta m)^2}}\ .
$$
The  radial deviation  $A(q)$ is then given by:
$$A(q)= v(q)-\tilde{v}(q)\ .$$
This deviation is plotted in FIG. \ref{fig:verzerrung} for three different
 masses in units of
the noncommutative scale (Planck mass):  straight line for  
$m=2.4\cdot 10^{-3}\cdot\lambda_{\rm nc}^{-1} $, 
dashed line for  $m=2.3\cdot 10^{-3}\cdot\lambda_{\rm nc}^{-1}$ and  
dashed-dotted line  for
$m=4.0\cdot10^{-3}\cdot\lambda_{\rm nc}^{-1} $; the momentum runs 
in the range  $q\in [0,0.008]\times \lambda_{\rm nc}^{-1}$ in units of the
 noncommutativity 
parameter $\lambda_{\rm nc}$.
\section{Discussion}
\noindent If we take a closer look at the connection between the physical
 mass and the bare mass 
in FIG. \ref{fig:mtilde} we find  that for particles with a physical mass of  
magnitude
ranging from GeV to TeV and even for particles with $\widetilde m =0$, 
we have a finite bare mass $m\geq m_{\rm c}>0$ 
of order $>10^{-3}\cdot\lambda_{\rm nc}$.
 If we plug in the Planck length, the critical bare mass is about
$m_{\rm c}\approx 2.8279\cdot 10^{16}$GeV and  thus 
(accidentally) at the mass scale of the Grand Unified Theories (GUTs). 

\ 

The dispersion relation  in FIG. \ref{fig:verzerrung} can be 
interpreted in two ways. First we consider a general undetermined 
noncommutativity 
parameter $\lambda_{\rm nc}$. Then a maximal deviation of  $A(q)\approx 0.03$
 in the group velocity 
of a particle at  momentum $q\approx 86$GeV and  a physical mass  
of $\widetilde m\approx 80$GeV  yields  a  bare mass $m\approx 128$GeV and a
noncommutativity parameter $\lambda_{\rm nc}^{-1}\approx 32$TeV, 
while a maximal deviation of $A(q)\approx 0.095$ at  momentum $q\approx 764$GeV
of a particle with physical mass $\widetilde m\approx 90$GeV thus a bare mass 
$m\approx 1309.1$GeV
leads to  $\lambda_{\rm nc}^{-1}\approx 545,5$TeV. 
Therefore  it should be possible to fix the energy  bounds of 
the noncommutativity parameter $\lambda_{\rm nc}$ from experiment.
\ 

We can also consider the second case were $\lambda_{\rm nc}\equiv\planckl$
 and particles
have physical masses of around 100GeV. For the dispersion relation we need to
know the bare mass, which in this case is of course larger than $m_{\rm c}$ 
(of order $>10^{16}$GeV). 
Then the deviation $A(q)$ is a curve somewhere between the dashed
and the straight line in FIG. \ref{fig:verzerrung} so that its local maximum 
is 
situated around  $q\approx 0.0012\cdot\planckl^{-1}$, which implies 
$q>10^{16}$GeV.
If we now consider momenta of the order GeV to TeV we will not be able to
detect any deviation in an experiment due to the vanishing of $A(q)$ in the 
limit $q\to 0$. 
We will see nothing of the noncommutative Minkowski space from this point of 
view.
So the ansatz of  the regularised,  ultraviolet finite Hamilton operator on 
the  noncommutative 
Minkowski space for regularising the UV- and the IR-regime does not
contradict any experiment if we assume the Planck scale as the 
noncommutativity scale.
\

We may compare this with  the Lorentzian invariant Yang-Feldman approach 
\cite{Bahns:2004fc} 
and the recent works \cite{Doescher:2006wp}
on the noncommutative Minkowski space, where a  deviation of the 
group velocity $v_\bot$ occurs (which in our approach is zero), which
 was shown to increase for decreasing masses.  
This is also true for our group velocity $v(q)$, but contrary to the Y-F-case,
 where the maximum 
is reached at zero momentum, our maximum is localised between  
$10^{-3}\cdotp\lambda_{\rm nc}$ and
 $\lambda_{\rm nc}$ and vanishes both for large and zero momentum 
(asymptotically free).
\ 

In \cite{Doescher:2006wp} it is concluded that it is improbable to see  
detectable  effects
at LHC of the noncommutative Minkowski space in the Y-F-approach in the case 
$\lambda_{\rm nc}\equiv\planckl$,
if one chooses the typical parameter of the Higgs field.
This is also the case in this approach due to the lower limit $m_{\rm c}$ of 
the bare mass
and the very small energy correction at momenta of magnitude GeV to TeV.

\begin{acknowledgments}
I would like to thank Klaus Fredenhagen for fruitful discussions and comments.
\end{acknowledgments}
\appendix*
\section{Energy correction in the vacuum state}
In this appendix the energy correction in the vacuum state is  calculated  as 
an example.
To keep the formulas short, the computations are restricted to a 
$\phi_R^{(3)}$-Theory; the general case
can be performed in analogy. 
The notation is adapted from \cite{Bahns:2003vb}.
Starting from equation \eqref{eq:Vakuumkorrektur1}:
\begin{eqnarray*}
\alpha_2[\Omega]=\frac{1}{3!^2}&&\int 
  \frac{d\pb_1 d\pb_2 d\pb_3}{\omega_{\pb_1}+\omega_{\pb_2}+\omega_{\pb_3}}
\Big[\\
&&\int d\xb d\yb  \lambda(\xb) \lambda(\yb) 
         \bra{\Omega}\wickprod{\phi_R^{(3)}(0,\xb)}\ket{p_1p_2p_3}\\
&&\quad\quad      \bra{p_1p_2p_3}\wickprod{\phi_R^{(3)}(0,\yb)} \ket{\Omega}
\Big]\ ,
\end{eqnarray*}
we  calculate first the following integral by standard methods:\\
$\int d\xb \lambda(\xb) \bra{\Omega}\wickprod{\phi_R^{(3)}(0,\xb)}
\ket{p_1p_2p_3}=$\\
\begin{eqnarray*}
 &=&\frac{c_3}{3^4}\frac{1}{\sqrt{3!}}\int d\avn_1\  
    \lambda(\kappab(\abn_1)) e^{-\frac{1}{2}|\abn_1-\kappab(\abn_1)|^2 } \  
                e^{-\frac{1}{2}|\avn^0_1|^2 }\\ 
 && \quad\times \quad \delta(\kappa^0(\avn_1^0))  
  \bra{\Omega}\wickprod{\phi(a_{11})\phi(a_{21})\phi(a_{31})}a^+_{\pb_1}
a^+_{\pb_2} a^+_{\pb_3}\ket{\Omega}\\
&& \ \\
&=& \frac{c_3}{3^4}\frac{1}{\sqrt{3!}} \frac{1}{(2\pi)^{9/2}}
\tfrac{1}{\sqrt{8}} \sum_\pi\int d\avn_1\ 
          \lambda(\kappab(\abn_1)) \\ 
  && \times\                e^{-\frac{1}{2}|\abn_1-\kappa(\abn_1)|^2 }
 e^{-\frac{1}{2}|\avn^0_1|^2 }  \delta(\kappa^0(\avn_1^0)) \Big\{\\
   &&\times\    \frac{e^{-ip_{\pi(1)}a_{11}}e^{-ip_{\pi(2)}a_{21}}
         e^{-ip_{\pi(3)}a_{31}}}{\sqrt{\omega_{\pb_1}\omega_{\pb_2}
\omega_{\pb_3}}}\Big\}\ \ .
\end{eqnarray*}
The sum runs over all permutations $\pi$ and in analogy we obtain: \\ 
$\int d\yb \lambda(\yb)\bra{p_1p_2p_3}\wickprod{\phi_R^{(3)}(0,\yb)}
\ket{\Omega}=$ \\
\begin{eqnarray*}
    &=& \frac{c_3}{3^4} \frac{1}{\sqrt{3!}}\int d\avn_2\ 
\lambda(\kappab(\abn_2))
 e^{-\frac{1}{2}|\abn_2-\kappab(\abn_2)|^2 }  
       e^{-\frac{1}{2}|\avn^0_2|^2 } \\ 
&&\ \times\   \delta(\kappa^0(\avn_2^0))
     \bra{\Omega}a_{\pb_1}a_{\pb_2}a_{\pb_3}
\wickprod{\phi(a_{12})\phi(a_{22})\phi(a_{32})}
         \ket{\Omega}\\
    &=&\frac{c_3}{3^4}\frac{1}{\sqrt{3!}}\frac{1}{(2\pi)^{9/2}}
\frac{1}{\sqrt{8}} \sum_\sigma 
\int d\avn_2\ \lambda(\kappab(\abn_2)) 
                  e^{-\frac{1}{2}|\abn_2-\kappab(\abn_2)|^2 }\\
    && \times\             e^{-\frac{1}{2}|\avn^0_2|^2 }  
\delta(\kappa^0(\avn_2^0))
       \frac{e^{ip_{\sigma(1)}a_{12}}e^{ip_{\sigma(2)}a_{22}}
          e^{ip_{\sigma(3)}a_{32}}}{\sqrt{\omega_{\pb_1}
\omega_{\pb_2}\omega_{\pb_3}}}\ .
\end{eqnarray*}
The following transformations are similar to the one in \cite{Bahns:2003vb} 
and render the Gaussian functions independent of one of the
integration variables. First of all we redefine the momenta:
\begin{eqnarray*}
U'_{1}\kvn_1:= -\left(\begin{matrix} I &0  & 0\\
                           0  &I & 0 \\ 
                           0  & 0  &I  \\
\end{matrix}\right)
\left(\begin{matrix}     p_{\pi(1)} \\
                        p_{\pi(2)}  \\ 
                        p_{\pi(3)}  \\
\end{matrix}\right)
\ ,
\end{eqnarray*}
\begin{eqnarray*}
U'_2\kvn_2:= \left(\begin{matrix} I &0  & 0\\
                                0  &I & 0 \\ 
                                0  & 0  &I  \\
\end{matrix}\right)
\left(\begin{matrix}      p_{\sigma(1)} \\
                          p_{\sigma(2)}  \\ 
                          p_{\sigma(3)}  \\
\end{matrix}\right)\ .
\end{eqnarray*}
These expressions substituted  into the upper integrals leads to the following:
\begin{eqnarray*}
\int {d\abn_1d\abn_2}\   \lambda(\kappab(\abn_1)) \lambda(\kappab(\abn_2)) 
                          e^{-\frac{1}{2}|\abn_1-\kappab(\abn_1)|^2 }
                         e^{-\frac{1}{2}|\abn_2-\kappab(\abn_2)|^2 }\ \\ 
\times    \quad\quad     e^{i\Ub_1\kbn_1\abn_1 } e^{i\Ub_2\kbn_2\abn_2 }\\ 
\times \int d\avn_1^0 d\avn_2^0\  e^{-\frac{1}{2}|\avn_1^0|^2}
e^{-\frac{1}{2}|\avn_2^0|^2}
 \delta(\kappa^0(\avn_1^0)) \delta(\kappa^0(\avn_2^0))\\ 
\times    \quad\quad   e^{iU_1\kvn_1^0\avn_1^0} e^{iU_2\kvn_2^0\avn_2^0}\ .
\end{eqnarray*}
The techniques also used in \cite[appendix]{Bahns:2003vb} yield Gaussian 
functions in $N(n-1)$ coordinates
and an integral over $N$ coordinates $\beta_M=(\beta^0_M,\betab_{M}):=
(b^0_{nM},\bb_{nM})$:
\begin{eqnarray*}
&& (2\pi)^6 \  e^{-\frac{1}{2}|(\Ib-\Pb)\Ub_1\kbn_1|^2 } 
e^{-\frac{1}{2}|(\Ib-\Pb)\Ub_2\kbn_2|^2 }\\
&&\times \int{d\betab_1 d\betab_2}\ \lambda(\betab_1/\sqrt{3})
\lambda(\betab_2/\sqrt{3})\\ 
&&   \times   \quad\quad  e^{i(\Vb_{\underline{e}_3}\Rb\Ub_1 
\kbn_1)\cdot\betab_1}
         e^{i(\Vb_{\underline{e}_3}\Rb\Ub_2 \kbn_2)\cdot\betab_2}\\
&&\times\ \int d\bvn_1^0 d\bvn_2^0\  e^{-\frac{1}{2}(|\bvn_{1}^0|^2)}
e^{-\frac{1}{2}(|\bvn_2^0|^2)}
 \delta\left(\tfrac{b_{31}^0}{\sqrt{3}}\right) 
\delta\left(\tfrac{b_{32}^0}{\sqrt{3}}\right)\ \\ 
&&   \times   \quad\quad  
 e^{i(RU_1\kvn_1^0)\bvn_1^0} e^{i(RU_2\kvn_2^0)\bvn_2^0}\ .
\end{eqnarray*}
Performing the time-like integration in the variables 
$b_{11}^0,\ b_{21}^0,\ b_{12}^0,
b_{22}^0$ and setting $\beta_i=b_{3i}$ we get:
\begin{eqnarray*}
&& (2\pi)^6 \  e^{-\frac{1}{2}|(\Ib-\Pb)\Ub_1\kbn_1|^2 }
 e^{-\frac{1}{2}|(\Ib-\Pb)\Ub_2\kbn_2|^2 }\\
&&\times\  (2\pi)^2 \int{d\betab_1 d\betab_2}\ \lambda(\betab_1/\sqrt{3})
\lambda(\betab_2/\sqrt{3}) 
                    e^{i(\Vb_{\underline{e}_3}\Rb\Ub_1 \kbn_1)\cdot\betab_1}\\
&&\times\  e^{i(\Vb_{\underline{e}_3}\Rb\Ub_2 \kbn_2)\cdot\betab_2} 
         e^{-\frac{1}{2}|(I-P)U_1\kvn_1^0|^2}
 e^{-\frac{1}{2}|(I-P)U_2\kvn_2^0|^2}\\
&&\times\ \int d\beta_1^0 d\beta_2^0\  e^{-\frac{1}{2}|\beta_1^0|^2} 
                                       e^{-\frac{1}{2}|\beta_2^0|^2}
                 \delta\left(\tfrac{\beta_{1}^0}{\sqrt{3}}\right) 
                  \delta\left(\tfrac{\beta_{2}^0}{\sqrt{3}}\right)\\ 
  &&\times\ e^{i(V_{\underline{e}_3}RU_1\kvn_1^0)\beta_1^0} 
            e^{i(V_{\underline{e}_3}RU_2\kvn_2^0)\beta_2^0}\ .
\end{eqnarray*}
The last two lines give just the number of the Wickpower $n$ (i.e. 3). We can
 not set 
$\lambda(\betab_i)\to 1$ in the second line since we would obtain the square 
of a delta function. 
So we use the relation
$V_{\underline{e}_n}RU_1\kvn_1^\mu=\xin\cdot (U_1\kvn^\mu)=
\tfrac{1}{\sqrt{n}}\sum_i^n u_{ii1}k_{i1}^\mu$
(see \cite{Bahns:2003vb}) and  obtain  the following expression 
($p^0=\omega_\pb$):
\begin{eqnarray*}
\tilde{c}\sum_{\pi,\sigma}&&\int 
                     \frac{d\pb_1 d\pb_2 d\pb_3\ }{(
                         \omega_{\pb_1}+\omega_{\pb_2}+\omega_{\pb_3})
                        {\omega_{\pb_1}\omega_{\pb_2}\omega_{\pb_3}}}\\
&&\times  \ |\check\lambda(\tfrac{1}{{3}}(\pb_1+\pb_2+\pb_3))|^2\\
&&\times  \ \prod_{j=1}^3 e^{-\frac{1}{2}|p_{\pi(j)}-
\frac{1}{3}\sum_{i=1}^3 p_{\pi(i)}|^2 } 
 e^{-\frac{1}{2}|p_{\sigma(j)}-\frac{1}{3}\sum_{i=1}^3 p_{\sigma(i)}|^2 }\ .
\end{eqnarray*}
The constant in front of the sum is 
$\tilde{c}:=\frac{c_3^2}{3!^2}\frac{1}{{3!}} 
\frac{1}{(2\pi)^{9}}\frac{3 (2\pi)^8}{3^8 8}$ and $c_n=n^2(2\pi)^{-2(n-1)}$.
For each permutation the summands have the same value. Therefore 
we obtain the following integral, which is not defined for 
$\lambda(\xb)\to 1$ since 
in Fourier space $|\check\lambda(\pb)|^2\to |\delta(\pb)|^2$ would tend to 
the square of a delta function:
\begin{eqnarray}\label{eq:VakuumskorrekturA}
\frac{1}{3^4}\frac{1}{(2\pi)^9}\frac{1}{16}&&\int  
                         \frac{d\pb_1 d\pb_2 d\pb_3\ }{(
                 \omega_{\pb_1}+\omega_{\pb_2}+\omega_{\pb_3})
                {\omega_{\pb_1}\omega_{\pb_2}\omega_{\pb_3}}}\\\nonumber
&&\times  \ |\check\lambda(\tfrac{1}{{3}}(\pb_1+\pb_2+\pb_3))|^2\\\nonumber
&&\times  \ \prod_{j=1}^3 e^{-|p_{j}-\frac{1}{3}\sum_{i=1}^3 p_{i}|^2 }\ . 
\\\nonumber
\end{eqnarray}
The problem of the occurrence of the square of a delta function is absent in 
the one particle term due to different momenta.
\ 

With this procedure one can in principle calculate the correction of the
 improper one particle state or
also multiparticle states  to higher orders and/or higher powers of the 
regularised 
Wick monomials.
\vfill


\end{document}